\documentclass[12pt]{article}
\pdfoutput = 1
\begin{document}
\title
{Remnant mass and entropy of black holes and modified uncertainty principle}

\author{
{\bf {\normalsize Abhijit Dutta}$^{a} $\thanks{dutta.abhijit87@gmail.com}}, 
{\bf {\normalsize Sunandan Gangopadhyay}$^{b,c,d}
$\thanks{sunandan.gangopadhyay@gmail.com, sunandan@iucaa.ernet.in}}\\
$^{a}${\normalsize Department of Physics, Adamas Institute of Technology, Barasat, Kolkata 700126, India}\\
$^{b}$ {\normalsize National Institute for Theoretical Physics, Stellenbosch University, South Africa}\\
$^{c}${\normalsize Department of Physics, West Bengal State University, Barasat, Kolkata 700126, India}\\
$^{d}${\normalsize Visiting Associate in Inter University Centre for Astronomy $\&$ Astrophysics,}\\
{\normalsize Pune, India}\\
[0.3cm]
}
\date{}

\maketitle

\begin{abstract}
{\noindent In this paper, we study the thermodynamics of black holes using a generalized uncertainty principle (GUP)
with a correction term linear order in the momentum uncertainty.
The mass-temperature relation and heat capacity are calculated from which critical and remnant masses are obtained. The results are exact and are found to be identical.
The entropy expression gives the famous area theorem upto leading order corrections from GUP. 
In particular, the linear order term in GUP leads to a $\sqrt{A}$ correction to the area theorem. 
Finally, the area theorem can be expressed in terms of a new variable termed as reduced horizon area only when the calculation is done to the next higher order correction from GUP.

}

\end{abstract}

\maketitle

\noindent The idea of existence of a minimal length arises naturally in 
quantum gravity theories in the form of effective minimal uncertainty in position. For example, in string theory \cite{Magg1}-\cite{Venez}, it is impossible to improve the spatial resolution below the characteristic length of the string which is expected to be close or equal to Planck length. Based on these arguments, the conventional Heisenberg uncertainty principle has been modified to the generalized uncertainty principle (GUP) \cite{koni}, \cite{magg2}. 
This idea, proposed first in \cite{mead}, has led recently to a considerable amount of study in various areas of physics.
For instance, the laws of black hole thermodynamics \cite{hawk1}-\cite{bek} has been investigated under this modification \cite{adler}-\cite{nico}, quantum gravity corrections are computed in quantum systems (such as particle in a box, Landau levels and simple harmonic oscillator) \cite{das2}-\cite{das6},
Planck scale corrections are obtained in the phenomena of superconductivity and quantum Hall effect \cite{saurya}
and its implications has been studied in cosmology \cite{das7}, \cite{das8}.

In this paper, we will find the thermodynamic properties of the Schwarzschild and Reissner-Nordstr\"{o}m (RN)
black holes using the following form of the GUP proposed in the literature \cite{das2}
\begin{eqnarray}
\delta x\delta p&\geq&\frac{\hbar}{2}\left\{1-\frac{2\alpha l_p}{\hbar}\langle p\rangle + \frac{\beta^2 l_p^2}{\hbar^2}\langle p^2\rangle\right\}\nonumber\\
&\geq&\frac{\hbar}{2}\left\{1+\left(\frac{\alpha l_{p}}{\hbar\sqrt{\langle p^2\rangle}}+\frac{\beta^2 l_{p}^2}{\hbar^2}\right)(\Delta p)^2 +\frac{\beta^2 l_{p}^2}{\hbar^2}\langle p\rangle^2 
-\frac{2\alpha l_{p}}{\hbar}\sqrt{\langle p^2\rangle}\right\}~;~\beta=2\alpha
\label{gup}
\end{eqnarray} 
where $l_p$ is the Planck length ($\sim 10^{-35}m$) and $\alpha, \beta$ are dimensionless constants. 
We calculate the critical mass (below which the thermodynamic quantities become ill-defined) and the remnant
mass (at which the radiation process stops) for these black holes. 
Our results are analytically exact in contrast to earlier results \cite{rb}
where approximations have been made in the computation of the critical and remnant masses. 
We finally compute the entropy and obtain the well known area theorem with GUP corrections.
The linear order term in momentum uncertainty leads to a $\sqrt{A}$ correction to the area theorem. 
For the Schwarzschild black hole, we extend the computation of the entropy to the next leading order correction
from the GUP and obtain the area theorem in terms of a new variable which can be interpreted as the reduced horizon area.
The reduced horizon area is found to have a singularity in the entropy which is avoided as the remnant mass is larger than the
mass for which the entropy becomes singular. However, the reduced horizon area 
does not come in the picture if we keep our computation to the leading order correction from GUP.

To start with, we first make a few remarks about the GUP (\ref{gup}). Note that in writing the second line of the above inequality, the relation $(\Delta p)^2 =\langle p^2\rangle-\langle p\rangle^2$ has been used. Now making the approximation
$\langle p\rangle\approx0$, the above relation can be put in the form
\begin{eqnarray}
\delta x\delta p&\geq&\frac{\hbar}{2}\left\{1-\frac{\alpha l_p}{\hbar}\Delta p 
+\frac{\beta^2 l_{p}^2}{\hbar^2}(\Delta p)^2\right\}.
\label{gup-1}
\end{eqnarray} 
However, to carry out our calculations. we shall consider $\alpha\neq\beta$ and at the end we shall put $\alpha=\beta/2$. 
This would allow us to check the consistency of our results with earlier findings \cite{abhi} by setting $\alpha=0$.

Let us now consider a Schwarzchild black hole of mass $M$. For any quantum particle (massless) near the horizon of the black hole, the momentum uncertainty can be written in terms of the temperature  as \cite{adler}
\begin{eqnarray}
T=\frac{(\delta p) c}{k_B}
\label{e2}
\end{eqnarray}
where $c$ is the speed of light and $k_B$ is the Boltzmann constant. 
At thermodynamic equilibrium, the temperature of the black hole will be equal to the temperature of the particle. 
Also, near the horizon of the Schwarzschild black hole, the position uncertainty of a particle will be of the order of the Schwarchild radius of the black hole \cite{adler},\cite{medved}
\begin{eqnarray}
\delta x=\epsilon r_{s}~;~
r_{s}=\frac{2GM}{c^2}
\label{e3}
\end{eqnarray}
where $\epsilon$ is a calibration factor, $r_s$ is the Schwarzschild radius and $G$ is the Newton's universal gravitational constant.

\noindent To relate the temperature with the mass of the black hole, the GUP (\ref{gup-1}) has to be saturated
\begin{eqnarray}
\delta x\delta p = \frac{\hbar}{2}\left\{1-\frac{\alpha l_p}{\hbar}(\delta p) + \frac{\beta^2 l_p^2}{\hbar^2}(\delta p)^2\right\}.
\label{e4}
\end{eqnarray}
This can be put in the following form using  eqs.(\ref{e2}) and (\ref{e3})
\begin{eqnarray}
M&=&\frac{M_{p}^2 c^2}{4\epsilon k_{B}T}\left\{1-\frac{\alpha l_p}{c \hbar}(k_{B}T)+ \frac{\beta^2 l_p^2}{c^2 \hbar^2}(k_{B}T)^2\right\}\nonumber\\
&=&\frac{M_{p}^2 c^2}{4\epsilon}\left\{\frac{1}{k_B T}-\alpha\frac{1}{M_p c^2}+\beta^2\frac{k_{B}T}{(M_p c^2)^2}\right\}
\label{e5}
\end{eqnarray}
where the relations $\frac{c\hbar}{l_p}=M_{p}c^{2}$ and $M_{p}=\frac{c^2 l_p}{G}$ ($M_{p}$ being the Planck mass) has been used.
 
\noindent In the absence of correction due to quantum gravity, eq.(\ref{e5}) reduces to
\begin{eqnarray}
M=\frac{M_{p}^2 c^2}{4\epsilon k_{B}T}~.
\label{e6}
\end{eqnarray}
Comparing this with the semi-classical Hawking temperature $T=\frac{M_{p}^{2}c^2}{8 \pi M k_{B}}$ \cite{hawk1}, \cite{hawk2}, yields the value of $\epsilon=2\pi$.

\noindent Hence, the mass-temperature  relation (\ref{e5}) takes the form
\begin{eqnarray}
M&=&\frac{M_{p}^2 c^2}{8\pi}\left\{\frac{1}{k_B T}-\alpha\frac{1}{M_p c^2}+\beta^2\frac{k_{B}T}{(M_p c^2)^2}\right\}.
\label{e7}
\end{eqnarray}
Now by definition, the heat capacity of the black hole reads
\begin{eqnarray}
C=c^2\frac{dM}{dT}
\label{e8}
\end{eqnarray}
which by using eq.(\ref{e7}) gives
\begin{eqnarray}
C=\frac{k_B}{8\pi}\left\{-\left(\frac{M_p c^2}{k_B T}\right)^{2} + \beta^2\right\}~.
\label{e9}
\end{eqnarray}
From a simple physical consideration (involving eqs.(\ref{e6}) and (\ref{e9})), one can argue that there exists a temperature at which the heat capacity vanishes \cite{abhi}. The radiation process stops at this temperature of the black hole with a finite mass termed as the remnant mass.

\noindent The entropy of the black hole can be determined from the first law of black hole thermodynamics as
\begin{eqnarray}
S=\int c^2 \frac{dM}{T}=\int C\frac{dT}{T}~.
\label{e10}
\end{eqnarray}
Using eq.(\ref{e9}) and computing the above integral yields
\begin{eqnarray}
S=\frac{k_B}{8 \pi}\left\{\frac{1}{2}\left(\frac{M_p c^2}{k_B T}\right)^2 
+ \beta^2 \ln\left(\frac{k_B T}{M_p c^2}\right)\right\}~.
\label{e11}
\end{eqnarray}
Now we will express $T$ in terms of $M$  so that the heat capacity and entropy can be represented in terms of $M$. To do that, we first introduce the following notations for convenience
\begin{eqnarray}
M'=\frac{8\pi M}{M_p}~;~T'=\frac{k_B T}{M_p c^2}~. 
\label{e12}
\end{eqnarray}
Eq.(\ref{e7}) can then be recast in the following form
\begin{eqnarray}
M'=\frac{1}{T'}-\alpha +\beta^2 T'
\label{e12s}
\end{eqnarray}
leading to the following quadratic equation in $T'$
\begin{eqnarray}
\beta^2 T'^2 -(\alpha+M') T' +1=0.
\label{e13}
\end{eqnarray}
Solving we get $T'(T)$ in terms of $M'(M)$ 
\begin{eqnarray}
T'=\frac{1}{2\beta^2}\left\{(\alpha+M')-\sqrt{(\alpha+M')^2-4\beta^2}\right\}
\label{e14}
\end{eqnarray}
where the negative sign before the square root has been taken to reproduce eq.(\ref{e6}) in the $\alpha, \beta\rightarrow0$ limit.
The above relation readily implies the existence of a critical mass below which the temperature will be a complex quantity 
\begin{eqnarray}
M_{cr}=\frac{(2\beta-\alpha)}{8\pi}M_p ~.
\label{e15}
\end{eqnarray}
In the limit $\alpha\rightarrow0$, the above result reduces to the one found earlier in \cite{abhi}. 
It is also evident that the linear term in the momentum uncertainty in eq.(\ref{gup-1}) reduces the critical mass. To get the actual critical mass, we set
$\alpha=\beta/2$, which yields
\begin{eqnarray}
M_{cr}=\frac{3\beta}{16\pi}M_p ~.
\label{e15z}
\end{eqnarray}
Now to find out the heat capacity in terms of $M'$, we square eq.(\ref{e14}) and substitute it in eq.(\ref{e9}) to get
\begin{eqnarray}
C=\frac{k_B}{8\pi}\left\{-\frac{2\beta^4}{(\alpha+M')^2 -2\beta^2 -(\alpha+M')\sqrt{(\alpha+M')^2 - 4\beta^2}} + \beta^2\right\}~.
\label{e16}
\end{eqnarray}
The remnant mass can now be obtained by setting $C=0$ (at which the radiation process stops) and this leads to
\begin{eqnarray}
M_{rem}=\frac{(2\beta-\alpha)}{8\pi}M_p ~.
\label{e17}
\end{eqnarray}
Note that the remnant and critical masses are equal as in \cite{abhi}. Once again we set $\alpha=\beta/2$, to get the actual remnant mass.

\noindent Finally, we would like to express the entropy in terms of the mass. To carry this out, we substitute eq.(\ref{e14}) in eq.(\ref{e11}) and carry out a binomial expansion keeping terms upto leading order $\alpha^2$, $\beta^2$ 
and $\alpha\beta^2$, to get
\begin{eqnarray}
\frac{S}{k_B} &=& \frac{4\pi M^2}{M_p^2}+ \alpha \frac{M}{M_p}+\frac{\alpha^2-2\beta^2}{16\pi} -\frac{\beta^2}{8\pi}\ln\left({\frac{8\pi M}{M_p}}\right)+\frac{\alpha \beta^2}{8\pi}\left(\frac{M_p}{8\pi M}\right)\nonumber\\
&=& \frac{S_{BH}}{k_B}+\frac{\alpha}{2\sqrt{\pi}}\sqrt{\frac{S_{BH}}{k_B}}+\frac{\alpha^2-2\beta^2}{16\pi}-\frac{\beta^2}{16\pi}\ln{\left(\frac{S_{BH}}{k_B}\right)}
-\frac{\beta^2}{16\pi}\ln{(16\pi)}\nonumber\\
&&+\frac{\alpha\beta^2}{32\pi^\frac{3}{2}}\sqrt{\frac{k_B}{S_{BH}}}
\label{e18}
\end{eqnarray}
where  $\frac{S_{BH}}{k_B} =\frac{4 \pi M^2}{M_p^2}$ is the semi-classical Bekenstein-Hawking entropy
for the Schwarzschild black hole.
  
\noindent  In terms of the area of the horizon $A=4\pi r_{s}^2 =16\pi \frac{G^2 M^2}{c^4} = 4 l_p^2 \frac{S_{BH}}{k_B}$,
eq.(\ref{e18}) can be recast in the following form
\begin{eqnarray}
\frac{S}{k_B} &=& \frac{A}{4 l_p^2}+\frac{\alpha}{2\sqrt{\pi}}\sqrt{\frac{A}{4 l_p^2}}+\frac{\alpha^2-2\beta^2}{16\pi}-\frac{\beta^2}{16\pi}\ln{\left(\frac{A}{4 l_p^2}\right)}- \frac{\beta^2}{16\pi}\ln{(16\pi)}
+\frac{\alpha\beta^2}{32\pi^\frac{3}{2}}\sqrt{\frac{4 l_p^2}{A}}\nonumber\\
\label{e19}
\end{eqnarray}
which in $\alpha\rightarrow0$ limit gives the same result as in \cite{abhi}. Note that in order to get the correct coefficients of the leading order correction terms, one has to set $\alpha=\beta/2$.
Interestingly, we find that the linear order term in GUP leads to a $\sqrt{A}$ correction to the area theorem. Such a term
has also been found in \cite{faragali} as a correction (due to GUP) to the bound of the maximal entropy of a bosonic field.
 
\noindent Keeping terms upto order $\beta^4$ and $\alpha^2 \beta^2$ in our calculation yields 
\begin{eqnarray}
\frac{S}{k_B} &=&\frac{A}{4 l_p^2}+\frac{\alpha}{2\sqrt{\pi}}\sqrt{\frac{A}{4 l_p^2}}+\frac{\alpha^2-2\beta^2}{16\pi}-\frac{\beta^2}{16\pi}\ln{\left(\frac{A}{4 l_p^2}\right)}- \frac{\beta^2}{16\pi}\ln{(16\pi)}+\frac{\alpha\beta^2}{32\pi^\frac{3}{2}}\sqrt{\frac{4 l_p^2}{A}}\nonumber\\&&+\frac{\beta^4}{64\pi^2}\frac{l_p^2}{A}- \frac{\alpha^2\beta^2}{256\pi^2}\left(\frac{4 l_p^2}{A}\right).
\label{e20}
\end{eqnarray}
Defining a new variable $A'$ as
\begin{eqnarray}
A' =A-\frac{l_p^2}{4\pi}(2\beta^2-\alpha^2)
\label{e21}
\end{eqnarray}
enables us to write eq.(\ref{e20}) as
\begin{eqnarray}
\frac{S}{k_B} &=& \frac{A'}{4l_p^2}+ \frac{\alpha}{2\sqrt{\pi}}\sqrt{\frac{A'}{4l_p^2}}-\frac{\beta^2}{16\pi}\ln{\left(\frac{A'}{4l_p^2}\right)}-\frac{\beta^2}{16\pi}\ln{(16\pi)}+\frac{\alpha\beta^2}{32\pi^\frac{3}{2}}\sqrt{\frac{4l_p^2}{A'}}-\frac{\beta^4}{256\pi^2}\left(\frac{4l_p^2}{A'}\right).\nonumber\\
\label{e22} 
\end{eqnarray}
The singular mass corresponding to zero reduced horizon area($A'$) can be easily computed and reads
\begin{eqnarray}
M_{sing} = \frac{\sqrt{2\beta^2-\alpha^2}}{8\pi}M_p .
\label{e23}
\end{eqnarray}
In the $\alpha\rightarrow0$ limit, we recover our earlier result \cite{abhi}. Setting $\alpha=\beta/2$, we get the actual singular mass to be
\begin{eqnarray}
M_{sing} = \frac{\sqrt{7}\beta}{16\pi}M_p .
\label{e23x}
\end{eqnarray}
Clearly the remnant mass (\ref{e17}) is greater than the singular mass which in turn implies that the black hole avoids approaching the singularity.
This completes our discussion of the effect of GUP on the thermodynamic properties of the Schwarzchild black hole.  
    
     
In the subsequent part, we consider the Reissner-Nordstr\"{o}m black hole of mass $M$ and charge $Q$ and study the effect of GUP on the thermodynamics of this black hole. 
In this case, the position uncertainty of a particle near the horizon of the black hole will be of the order of the RN radius of the black hole
\begin{eqnarray}
\delta x&=&\epsilon r_h\nonumber\\
r_h &=& \frac{Gr_0}{c^2}\nonumber\\
r_0 &=& M+\sqrt{M^2 -Q^2}
\label{e24}
\end{eqnarray}
where $r_h$ is the radius of the horizon of the RN black hole. The momentum uncertainty for this black hole 
can once again be written
in terms of the temperature as in eq.(\ref{e2}).
Following the analysis in \cite{abhi} using eq.(\ref{gup-1}), we get the relation between the mass, charge and temperature of this black hole to be 
\begin{eqnarray}
\frac{r_0^3}{Mr_0-Q^2} = \frac{M_{p}^2 c^2}{2 \pi}\left\{\frac{1}{k_B T}-\frac{\alpha}{M_p c^2}+\beta^2\frac{k_{B}T}{(M_p c^2)^2}\right\}.
\label{e28}
\end{eqnarray}
Now using the identity 
\begin{eqnarray}
\frac{r_{0}}{(Mr_0 -Q^2)}=\frac{1}{(r_0 -M)}
\label{e29}
\end{eqnarray}
eq.(\ref{e28}) can be recast as 
\begin{eqnarray}
\frac{r_{0}^2}{(r_0 -M)}=\frac{M_{p}}{2\pi}\left\{\frac{M_p c^2}{k_B T}-\alpha+\beta^2\frac{k_{B}T}{(M_p c^2)}\right\}.
\label{e30}
\end{eqnarray}
The heat capacity of the black hole can now be calculated using relation (\ref{e8}) and eq.(\ref{e30}) and becomes
\begin{eqnarray}
C=\frac{k_B (r_0 -M)^3}{2\pi r_{0}^2(2r_0 -3M)}\left\{-\left(\frac{M_p c^2}{k_B T}\right)^2 +\beta^2\right\}~.
\label{e31}
\end{eqnarray}
Now to express the heat capacity in terms of the mass, we once again use eq.(\ref{e12}) to put eq.(\ref{e30}) in the form
\begin{eqnarray}
\beta^{2} T'^{2} -h(r_0)T' +1=0\\
h(r_0)=\alpha+\frac{2\pi r_{0}^2}{M_p(r_0 -M)}~.\nonumber
\label{e32}
\end{eqnarray}    
Solving this equation, we get the expression for $T'(T)$ in terms of the mass and charge of RN black hole
\begin{eqnarray}
T'=\frac{h(r_0)}{2\beta^2}\left\{1-\sqrt{1-\frac{4\beta^2 }{[h(r_0)]^2}}\right\}~.
\label{e34}
\end{eqnarray}
The negative sign has been taken again before the square root to reproduce eq.(\ref{e14}) in the $Q\rightarrow0$ limit.
The above relation immediately gives the following condition for the temperature $T'(T)$ to be real
\begin{eqnarray}
1-\frac{4\beta^2 }{[h(r_0)]^2}\geq0~.
\label{e35}
\end{eqnarray}
Taking the equality sign in this condition finally gives the following cubic equation for the critical mass below which the temperature 
becomes a complex quantity
\begin{eqnarray}
4bM_{cr}^3 -b^2 M_{cr}^2 -4b M_{cr}Q^2 +Q^4 +b^2 Q^2=0~;~b=\frac{(2\beta-\alpha) M_p}{2\pi}~.
\label{e36}
\end{eqnarray}
Solving the above equation we get the expression for critical mass as
\begin{eqnarray}
M_{cr} = \frac{b}{12}\left[1+\frac{b^2+48Q^2}{B^{\frac{1}{3}}}+\frac{B^{\frac{1}{3}}}{b^2}\right]
\label{e37}
\end{eqnarray}
where
\begin{eqnarray}
B = b^6-144b^4Q^2-216b^2Q^4+12\sqrt{3}b^2\sqrt{-b^6Q^2+31b^4Q^4-112b^2Q^6+108Q^8}.
\label{e38}
\end{eqnarray}
It is easy to see that this reduces to the critical mass for the Schwarzschild black hole (\ref{e15})
in the $Q\rightarrow0$ limit, also in the $\alpha\rightarrow0$ limit reproduce the value for critical mass in \cite{abhi}.
To get the actual critical mass, we need to set $\alpha=\beta/2$ in the expression for $b$.
   
\noindent The heat capacity in terms of the mass can be obtained by squaring eq.(\ref{e34}) and substituting the expression in eq.(\ref{e31}): 
\begin{eqnarray}
C=\frac{k_B (r_0 -M)^3}{2\pi r_{0}^2(2r_0 -3M)}\left[-\frac{4\beta^4 }{[h(r_0)]^2}\frac{1}{\left\{1-\sqrt{1-\frac{4\beta^2 }{[h(r_0)]^2}}\right\}^2}+\beta^2\right] ~.
\label{e39}
\end{eqnarray}
To get the remnant mass, once again we set $C=0$ which gives the same cubic equation 
for the remnant as that for the critical mass (\ref{e36}). Therefore, the remnant mass is once again equal to the critical mass (for the RN black hole) 
similar to the Schwarzschild black hole.

\noindent Finally, we compute the entropy of the RN black hole. 
Using the expressions for temperature (\ref{e34}) and substituting this in the expression for entropy (\ref{e10}), we get
\begin{eqnarray}
\frac{S}{k_{B}}&=&\frac{2\beta^2}{M_p} \int \frac{dM}{g_1\left\{1-\sqrt{1-\frac{4\beta^2}{g_1}}\right\}}\nonumber\\
&=&\frac{\pi r_{h}^2}{l_{p}^2}+\frac{\alpha}{2}\frac{r_h}{l_p}-\frac{\beta^2}{16\pi}\ln\left(\frac{\pi r_{h}^2}{l_{p}^2}\right)
-\frac{\beta^2 Q^2}{8M_p^2 \left(\frac{\pi r_{h}^2}{l_{p}^2}\right)}\left\{1-\frac{\pi Q^2}{4M_{p}^2\left(\frac{\pi r_{h}^2}{l_{p}^2}\right)}\right\}\nonumber\\&&-\frac{\alpha\beta^2}{64}\left[-\frac{2}{\pi^\frac{3}{2}\sqrt{\left(\frac{\pi r_{h}^2}{l_{p}^2}\right)}}+\frac{2Q^2}{M_p^2\pi^\frac{1}{2}\left(\frac{\pi r_{h}^2}{l_{p}^2}\right)^\frac{3}{2}}-\frac{6Q^4\sqrt{\pi}}{5M_p^4\left(\frac{\pi r_{h}^2}{l_{p}^2}\right)^\frac{5}{2}}+\frac{2Q^6\pi^\frac{3}{2}}{7M_p^6\left(\frac{\pi r_{h}^2}{l_{p}^2}\right)^\frac{7}{2}}\right]\nonumber\\
&=&\frac{S_{BH}}{k_B}+\frac{\alpha}{2\sqrt{\pi}}\sqrt\frac{S_{BH}}{k_B}-\frac{\beta^2}{16\pi}\ln\left(\frac{S_{BH}}{k_B}\right)
 -\frac{\beta^2 Q^2}{8M_p^2 \left(\frac{S_{BH}}{k_B}\right)}\left\{1-\frac{\pi Q^2}{4M_{p}^2\left(\frac{S_{BH}}{k_B}\right)}\right\}\nonumber\\&&-\frac{\alpha\beta^2}{64}\left[-\frac{2}{\pi^\frac{3}{2}\sqrt{\left(\frac{S_{BH}}{k_B}\right)}}+\frac{2Q^2}{M_p^2\pi^\frac{1}{2}\left(\frac{S_{BH}}{k_B}\right)^\frac{3}{2}}-\frac{6Q^4\sqrt{\pi}}{5M_p^4\left(\frac{S_{BH}}{k_B}\right)^\frac{5}{2}}+\frac{2Q^6\pi^\frac{3}{2}}{7M_p^6\left(\frac{S_{BH}}{k_B}\right)^\frac{7}{2}}\right]
\label{e40}
\end{eqnarray}
where  $\frac{S_{BH}}{k_B} =\frac{\pi r_{h}^2}{l_{p}^2}$ is the semi-classical  Bekenstein-Hawking entropy for the RN black hole.

\noindent  In terms of the area of the horizon $A =4\pi r_{h}^2 =4l_{p}^2 \frac{S_{BH}}{k_B}$,
the above equation can be recast as
\begin{eqnarray}
\frac{S}{k_B} &=&\frac{A}{4l_p^2}+ \frac{\alpha}{2\sqrt{\pi}}\sqrt\frac{A}{4l_p^2} -\frac{\beta^2}{16\pi}\ln\left(\frac{A}{4l_p^2}\right) -\frac{\beta^2 Q^2}{8M_p^2 \left(\frac{A}{4l_p^2}\right)}\left\{1-\frac{\pi Q^2}{4M_{p}^2\left(\frac{A}{4l_p^2}\right)}\right\} \nonumber\\&&-\frac{\alpha\beta^2}{64}\left[-\frac{2}{\pi^\frac{3}{2}\sqrt{\left(\frac{A}{4l_p^2}\right)}}+\frac{2Q^2}{M_p^2\pi^\frac{1}{2}\left(\frac{A}{4l_p^2}\right)^\frac{3}{2}}-\frac{6Q^4\sqrt{\pi}}{5M_p^4\left(\frac{A}{4l_p^2}\right)^\frac{5}{2}}+\frac{2Q^6\pi^\frac{3}{2}}{7M_p^6\left(\frac{A}{4l_p^2}\right)^\frac{7}{2}}\right]
\end{eqnarray}
which is the area theorem with GUP corrections for the RN black hole.

We conclude by summarizing our findings. In this paper, 
we study the effect of GUP (with a linear order term in momentum uncertainty) in the thermodynamics of Schwarzschild and Reissner-Nordstr\"{o}m black holes.
We compute the critical and remnant masses for these black holes and observe that both get reduced in the presence of the linear order term.
The computations are exact in contrast to those done in \cite{rb}. 
These results may have possible applications in explaining the formation of black holes at energies
higher than the energy scales of LHC \cite{lhc1}-\cite{alifar}.
The expression for the entropy reveals the famous area theorem upto corrections terms from GUP. The
leading order correction contains a $\sqrt{A}$ term arising due to the presence of a linear order term in momentum uncertainty in GUP. Inclusion of higher order corrections 
reveal that the corrected area theorem can be rewritten in terms of a new variable  which can be termed as reduced horizon area. The entropy shows a singularity at zero reduced
horizon area from which the singular mass is obtained. However, the black hole avoids the singularity since the remnant mass (at which the black hole stops radiating) is greater than the singular mass.


\section*{Acknowledgements}The authors would like to thank the referee for useful comments.



\begin{thebibliography}{99}

\bibitem{Magg1}M. Maggiore, Phys. Lett. B 319, 83 (1993)
\bibitem{Venez}G. Veneziano, D. Amati, and M. Ciafaloni, Phys. Lett. B
216, 41 (1989).
\bibitem{koni}K. Konishi, G. Paffuti and P. Provero, Minimum physical length and the generalized uncertainty principle in string theory, Phys. Lett. B 234 (1990) 276.
\bibitem{magg2}M. Maggiore, The algebraic structure of the generalized uncertainty principle, Phys. Lett. B
319 (1993) 83 [hep-th/9309034]; Quantum groups, gravity, and the generalized uncertainty principle, Phys. Rev. D 49 (1994) 5182 [hep-th/9305163]
\bibitem{mead}C.A.~Mead, Phys. Rev. D 135 (1964) 849
\bibitem{hawk1}S.W.~Hawking, Nature (London) 248 (1974) 30.
\bibitem{hawk2}S.W.~Hawking, Commun. Math. Phys. 43 (1975) 199.
\bibitem{bek} J.D.~Bekenstein, Phys. Rev. D 7 (8) (1973) 2333
\bibitem{adler}R.J.~ Adler, P.~Chen, D.I.~ Santiago, Gen. Rel. Grav. 33 (2001) 2101; [gr-qc/0106080].
\bibitem{barun}B.~Majumder, Phys. Lett. B 703 (2011) 402.
\bibitem{rb}R.~Banerjee, S.~Ghosh, Phys. Lett. B 688 (2010) 224; arXiv:1002.2302 [gr-qc].
\bibitem{nico}M.~Isi, J.~Mureika, P.~Nicolini, JHEP 1311(2013) 139; arXiv:1310.8153 [hep-th].
\bibitem{das2}S.~Das, E.C.~Vagenas, Phys. Rev. Lett. 101 (2008) 221301; arXiv:0810.5333 [hep-th].
\bibitem{das3}S.~Das, E.C.~Vagenas, Phys. Rev. Lett. 104 (2010) 119002; arXiv:1003.3208 [hep-th].
\bibitem{das4}S.~Das, E.C.~Vagenas, Can. J. Phys. 87 (2009) 233; arXiv:0901.1768 [hep-th].
\bibitem{das5}S.~Das, E.C.~Vagenas, A.F.~Ali, Phys. Lett. B 690 (2010) 407, Erratum-ibid. 692 (2010) 342; 
arXiv:1005.3368 [hep-th].
\bibitem{das6}A.F.~Ali, S.~Das, E.C.~Vagenas,  Phys. Rev. D 84 (2011) 044013; arXiv:1107.3164 [hep-th].
\bibitem{saurya}S.~Das, R.B.~Mann, Phys. Lett. B 704 (2011) 596–599.
\bibitem{das7}S.~Basilakos, S.~Das, E.C.~Vagenas, JCAP 1009 (2010) 027; arXiv:1009.0365 [hep-th].
\bibitem{das8}W.~Chemissany, S.~Das, A.F.~Ali, E.C.~Vagenas, JCAP 1112 (2011) 017; arXiv:1111.7288 [hep-th].
\bibitem{abhi}S. Gangopadhyay, A.  Dutta, A. Saha, Gen.Rel.Grav. 46 (2014) 1661; arXiv:1307.7045[hep-th].
\bibitem{medved}A.J.M.~Medved, E.C.~Vagenas, Phys. Rev. D 70 (2004) 124021.
\bibitem{faragali}A.F.~Ali, Class.Quant.Grav. 28 (2011) 065013; arXiv:1101.4181 [hep-th].
\bibitem{lhc1}CMS collaboration, V. Khachatryan et al., Search for microscopic black hole signatures at
the Large Hadron Collider, Phys. Lett. B 697 (2011) 434.
\bibitem{lhc2}CMS collaboration, S. Chatrchyan et al., Search for microscopic black holes in pp collisions
at $\sqrt{s}= 7$ TeV, JHEP 04 (2012) 061.
\bibitem{alifar}A.F.~Ali, JHEP 1209 (2012) 067; 1208.6584 [hep-th].


\end{thebibliography}
 \end{document}